\documentclass[12pt]{article}
\usepackage{epsf,a4,latexsym}
\usepackage[small,bf]{caption}

\newcommand{\half}{\mbox{\small $\frac{1}{2}$}}          
\newcommand{\third}{\mbox{\small $\frac{1}{3}$}}         
\newcommand{\threehalf}{\mbox{\small $\frac{3}{2}$}}     
\newcommand{\twonineth}{\mbox{\small $\frac{2}{9}$}}     
\newcommand{\R}{\mbox{\tiny $R$}}                        
\newcommand{\Si}{\mbox{\tiny $S$}}                       
\newcommand{\NS}{\mbox{\tiny $N\!S$}}                    

\def\lsim{\mathrel{\rlap{\lower4pt\hbox{\hskip1pt$\sim$}}
    \raise1pt\hbox{$<$}}}                
\def\gsim{\mathrel{\rlap{\lower4pt\hbox{\hskip1pt$\sim$}}
    \raise1pt\hbox{$>$}}}                
\begin{document}

\title{
\vspace{-3.0cm}
\flushright{\normalsize DESY 10-029} \\
\vspace{-0.35cm}
{\normalsize Edinburgh 2010/06} \\
\vspace{-0.35cm}
{\normalsize LTH 867} \\
\vspace{-0.35cm}
{\normalsize March 2010} \\
\vspace{0.5cm}
\centering{\Large \bf Tuning the strange quark mass in lattice simulations}}

\author{\large
        W. Bietenholz$^a$, V. Bornyakov$^b$,
        N.~Cundy$^c$, M.~G\"ockeler$^c$, \\
        R.~Horsley$^d$, A.~D. Kennedy$^d$,
        W.~G. Lockhart$^e$, Y.~Nakamura$^c$, \\
        H.~Perlt$^f$, D.~Pleiter$^g$,
        P.~E.~L.~Rakow$^e$, A.~Sch\"afer$^c$, \\
        G.~Schierholz$^{ch}$, A.~Schiller$^f$,
        H.~St\"uben$^i$ and J.~M.~Zanotti$^d$ \\[0.3em]
         -- QCDSF-UKQCD Collaboration -- \\[1em]
        \small $^a$ Instituto de Ciencias Nucleares,
               Universidad Aut\'{o}noma de M\'{e}xico,\\[-0.5em]
        \small A.P. 70-543, C.P. 04510 Distrito Federal, Mexico \\[0.25em]
        \small $^b$ Institute for High Energy Physics,
               142281 Protovino, Russia and \\[-0.5em]
        \small Institute of Theoretical and
               Experimental Physics,
               117259 Moscow, Russia \\[0.25em]
        \small $^c$ Institut f\"ur Theoretische Physik,
               Universit\"at Regensburg, \\[-0.5em]
        \small 93040 Regensburg, Germany \\[0.25em]
        \small $^d$ School of Physics and Astronomy,
               University of Edinburgh, \\[-0.5em]
        \small Edinburgh EH9 3JZ, UK \\[0.25em]
        \small $^e$ Theoretical Physics Division,
               Department of Mathematical Sciences, \\[-0.5em]
        \small University of Liverpool,
               Liverpool L69 3BX, UK \\[0.25em]
        \small $^f$ Institut f\"ur Theoretische Physik,
               Universit\"at Leipzig, \\[-0.5em]
        \small 04109 Leipzig, Germany \\[0.25em]
        \small $^g$ John von Neumann Institute NIC / DESY Zeuthen, \\[-0.5em]
        \small 15738 Zeuthen, Germany \\[0.25em]
        \small $^h$ Deutsches Elektronen-Synchrotron DESY, \\[-0.5em]
        \small 22603 Hamburg, Germany \\[0.25em]
        \small $^i$ Konrad-Zuse-Zentrum
               f\"ur Informationstechnik Berlin, \\[-0.5em]
        \small 14195 Berlin, Germany }

\date{March 4, 2010}

\maketitle


\clearpage

\begin{abstract}
   QCD lattice simulations with $2+1$ flavours typically start at
   rather large up-down and strange quark masses and extrapolate
   first the strange quark mass to its physical value and then
   the up-down quark mass. An alternative method of tuning the
   quark masses is discussed here in which
   the singlet quark mass is kept fixed, which ensures that the
   kaon always has mass less than the physical kaon mass.
   It can also take into account the different renormalisations
   (for singlet and non-singlet quark masses) occurring for non-chirally
   invariant lattice fermions and so allows a smooth extrapolation
   to the physical quark masses.
   This procedure enables a wide range of quark masses to be probed,
   including the case with a heavy up-down quark mass and light
   strange quark mass. Results show the correct order
   for the baryon octet and decuplet spectrum and an extrapolation
   to the physical pion mass gives mass values to within a few percent
   of their experimental values.
\end{abstract}

\clearpage


\section{Introduction}


There has been a steady progression of lattice results from a quenched sea
to a two-flavour and more recently $2+1$ flavour sea in an attempt
to provide a more complete and quantitative description of hadronic phenomena.
(By $2+1$ flavours we mean here two mass degenerate up-down, $m_l^{\R}$,
quarks and one strange, $m_s^{\R}$, quark.) In this letter we discuss
some ways of approaching in the $m_l^{\R}$--$m_s^{\R}$ plane the physical point
$(m_l^{\R*}, m_s^{\R*})$, where the natural starting point for these paths
is an $SU(3)$ flavour symmetric point $m_l^{\R} = m_s^{\R} = m_{sym}^{\R(0)}$.
(The superscripts ${}^*$, ${}^{(0)}$ denote the physical point
and flavour symmetric point respectively and ${}^{\R}$ means the 
renormalised quantity.) The usual procedure is to estimate
the physical strange quark mass and then try to keep it fixed,
i.e.\ $m^{\R}_s = \mbox{constant}$, as the light quark mass is reduced
to its physical value. However the problem is that the kaon mass
is always larger than its physical value. We propose here
instead to choose the path such that the singlet quark mass is kept fixed,
\begin{equation}
   \overline{m}^{\R} = \third ( 2m^{\R}_l + m^{\R}_s ) = \mbox{constant}\,.
\end{equation}
This procedure has the advantage that we can vary both quark masses
over a wide range, and is thus particularly useful for strange
quark physics. $SU(3)_F$ chiral perturbation theory should work well,
because both the kaon and $\eta$ are lighter than their physical values
along the entire trajectory. (They both approach their final mass values
from below.) Since $SU(3)_F$ chiral perturbation theory is thought
to be valid for $m_K < 600\,\mbox{MeV}$, \cite{leutwyler}, we should
always be able to make use of chiral perturbation theory. If we extend
our measurements beyond the symmetric point we can also investigate a world
with heavy up-down quarks and a lighter strange quark.

As a `proof of concept' results given here show firstly the correct order
for the baryon octet and decuplet mass spectrum and secondly an
extrapolation to the physical pion mass yields results to
within a few percent for the baryon masses.


\section{Extrapolating flavour singlet quantities}


Flavour singlet quantities are flat at a point on the $SU(3)$
flavour symmetric line and hence allow simpler extrapolations
to the physical point. This may be shown by considering small
changes about a point on the flavour symmetric line.
Let $X_S(m_u^{\R}, m_d^{\R}, m_s^{\R})$ be a flavour singlet object
i.e.\ $X_S$ is invariant under the quark permutation symmetry
between $u$, $d$ and $s$. So Taylor expanding $X_S$ about a point
on the symmetric line where flavour $SU(3)$ holds gives
\begin{eqnarray}
   \lefteqn{
      X_S( \overline{m}^{\R(0)} + \delta m_l^{\R},
           \overline{m}^{\R(0)} + \delta m_l^{\R},
           \overline{m}^{\R(0)} + \delta m_s^{\R}) }
     & &                                                     \nonumber \\
     &=& X_{S\,sym}^{(0)}
      + \left. { \partial X_S \over \partial m^{\R}_u } \right|^{(0)}_{sym} \,
                                         \hspace*{-0.20in}\delta m_l^{\R}
      + \left. { \partial X_S \over \partial m^{\R}_d } \right|^{(0)}_{sym} \,
                                         \hspace*{-0.20in}\delta m_l^{\R}
      + \left. { \partial X_S \over \partial m^{\R}_s } \right|^{(0)}_{sym} \,
                                         \hspace*{-0.20in}\delta m_s^{\R} 
      + O( (\delta m_q^{\R})^2 ) \,.
\end{eqnarray}
But on the symmetric line we have
\begin{equation}
   \left. { \partial X_S \over \partial m^{\R}_u } \right|_{sym}
      = \left. { \partial X_S \over \partial m^{\R}_d } \right|_{sym}
      = \left. { \partial X_S \over \partial m^{\R}_s } \right|_{sym} \,,
\end{equation}
and on our chosen trajectory $\overline{m}^{\R} = \mbox{constant}$,
\begin{equation}
   2\delta m_l^{\R} + \delta m_s^{\R} = 0 \,,
\end{equation}
which together imply that
\begin{eqnarray}
   X_S( \overline{m}^{\R(0)} + \delta m_l^{\R},
        \overline{m}^{\R(0)} + \delta m_l^{\R},
        \overline{m}^{\R(0)} + \delta m_s^{\R})
     = X_{S\,sym}^{(0)} + O( (\delta m_q^{\R})^2 ) \,.
\label{Xs_expansion}
\end{eqnarray}
In other words, the effect at first order of changing the
strange quark mass is cancelled by the change in the light quark mass,
so we know that $X_S$ must have a stationary point on the $SU(3)_F$
symmetric line. If we were making a quadratic extrapolation to the
physical point, this eliminates a free parameter -- we only need two
parameters to make the quadratic extrapolation. This is particularly useful
for quantities like the force scale, $r_0$, where we do not have
any theoretical input from chiral perturbation theory.
Also the fact that $X_S$ is flat at the symmetric point means
that the extrapolated value cannot lie very far from the measured value.
Other paths are not so fortunate. What we are doing is to keep the
flavour singlet part of the quark matrix constant, while increasing
the octet part (the changes in quark masses are proportional to $\lambda_8$).
Gluons are flavour-blind, they cannot couple directly to a
flavour octet operator, they can only see flavour singlets --
which can only occur in the square of the octet part of the mass.
So everything about our gluon configuration will vary
quadratically with the distance from the symmetric point.
So we are generating all our configurations closer to the physical
point by taking the $\overline{m}^{\R} = \mbox{constant}$ line.

Other potential advantages include:
as $m^{\R}_l \searrow m^{\R*}_l$ then $m^{\R}_s \nearrow m^{\R*}_s$, i.e.\ ,
the $m^{\R}_s$--$m^{\R}_l$ splitting or $m_K$ increases to its physical value.
The singlet quark mass is correct from the very beginning;
numerically the simulation cost change should be moderate and the
update algorithm is expected to equilibrate quickly along this path.

For $X_S$ we have several possibilities: for example the centre of
mass squared of the meson octet, $\third(m_\pi^2 + 2m_K^2)$
or the centre of mass of the baryon octet or decuplet,
$X_N = \third(m_N+m_\Sigma+m_\Xi)= 1.150\,\mbox{GeV}$,
$X_\Delta = \third(2m_\Delta+m_\Omega)= 1.379\,\mbox{GeV}$
respectively or a gluonic quantity such as $X_r = 1/r_0$.

We can check the above result, eq.~(\ref{Xs_expansion}), by considering
leading order (LO) together with next to leading order (NLO) $SU(3)_F$
chiral perturbation theory, $\chi$PT. Note that LO $\chi$PT corresponds
to linear terms in the expansion of $X_S$ and so from eq.~(\ref{Xs_expansion})
will be absent. Rather than using the full symmetric $1+1+1$ results
it is sufficient to just consider the $2+1$ results. Let us first
define $\chi_q = 2B_q^{\R}m_q^{\R}$ and furthermore set
$\chi_\eta = (\chi_l + 2\chi_s)/3$ and $\chi_K = (\chi_l + \chi_s)/2$.
Then $\chi_l + \chi_\eta = 2\overline{\chi}$ and
$\chi_l + 2\chi_K =3\overline{\chi}$ are constants on
our trajectory and so 
\begin{eqnarray}
   \delta\chi_l + \delta\chi_\eta = 0 = \delta\chi_l + 2\delta\chi_K \,.
\end{eqnarray}
This means that any functions of the form
\begin{eqnarray}
   f(\chi_\eta) + f(\chi_l)
   \quad \mbox{or} \quad
   2h(\chi_K) + h(\chi_l) \,,
\end{eqnarray}
will have zero derivative at the symmetric point, so they are
permitted as higher order corrections. Using the LO and NLO results from
e.g.\ \cite{allton08a,walker-loud04a,tiburzi04a} (where further details
of the functions can be found) we have
\begin{eqnarray}
   \third ( 2m_K^2 + m_\pi^2)
      &=& \overline{\chi} + \{ f_\pi(\chi_\eta) + f_\pi(\chi_l) \}
                                                            \nonumber \\
   \third ( m_N + m_\Sigma + m_\Xi )
      &=& m_{0N} + 2(\alpha_N + \beta_N +3\sigma_N) \overline{\chi} 
                                                            \nonumber \\
      & & +  \{ f_N(\chi_\eta) + f_N(\chi_l)
                + 2h_N(\chi_K) + h_N(\chi_l) \}
                                                            \nonumber \\
   \third ( 2m_\Delta + m_\Omega )
      &=& m_{0\Delta} + 2(\gamma_\Delta -3\sigma_\Delta) \overline{\chi}
                                                            \nonumber \\
      & &  + \{ f_\Delta(\chi_\eta) + f_\Delta(\chi_l)
                + 2h_\Delta(\chi_K) + h_\Delta(\chi_l) \} \,,
\end{eqnarray}
with 
\begin{equation}
   f_\pi(\chi)
      = \alpha_\pi \overline{\chi}^2 
          + \beta_\pi \chi^2 + \gamma_\pi \ln (\chi / \Lambda_\chi) \,,
\end{equation}
and
\begin{eqnarray}
   f_S(\chi)
      &=& \delta_S\chi^{\threehalf} + \epsilon_S F(\chi^{\half})
                                                            \nonumber \\
   h_S(\chi)
      &=& \zeta_S\chi^{\threehalf} + \eta_S F(\chi^{\half}) \,,
\end{eqnarray}
(with $S = N$, $\Delta$). $\alpha$, $\ldots$, $\eta$
are combinations of the low energy constants. The NLO results are
shown in curly brackets. These results are thus all in agreement
with our previous discussion. Note also that on the
$\overline{m}^{\R} = \mbox{constant}$ trajectory some of the low energy
constants combine together leading to fits with fewer free parameters. 

We now have to relate the known physical point to the initial
symmetric point. As discussed above, we expect $X_S$
to be constant (in $m_l^{\R}$) up to small corrections, so we will find it
sufficient to consider only LO $\chi$PT. Then keeping $\overline{m}^{\R}$
constant means keeping $(2m_K^2 + m_\pi^2)/3$ constant. For this
path choice, we can now relate the known physical point to the
initial symmetric point,
\begin{equation}
   \left. { {1 \over 3} ( 2m_K^2 + m_\pi^2 ) \over X_S^2 } \right|^*
          = \left. { m_\pi^2 \over X_S^2 } \right|^{(0)}_{sym} \,,
\label{physical_point}
\end{equation}
where $S = N$, $\Delta$ and $r$ respectively. So simulations
along the flavour symmetric line and using eq.~(\ref{physical_point})
are sufficient to determine the initial point. This procedure is
illustrated in Fig.~\ref{b5p50_mps2oX2_2mpsK2-mps2oX2} in the next section.

If we now generalise to consider higher order terms from the above
discussion lines of constant $X_S$ are curved though they do still have 
to have the slope of $-2$ at the point where they cross the
$m_l^{\R} = m_s^{\R} \equiv m_{sym}^{\R}$ line. Now we have to
specify more closely what we mean when we keep $2m_K^2 + m_\pi^2$
constant, as different scale choices give different paths.
As shown above in eq.~(\ref{Xs_expansion}) on our trajectory
this is a higher order effect, on other choices of trajectory
the difference between different scale definitions would
be much more important. If we make different choices of the quantity we
keep constant at the experimentally measured physical value, for
example the choices discussed above we get slightly different
trajectories. The different trajectories begin at slightly different
points along the symmetric line. Initially they are all parallel
with slope $-2$, but away from the symmetry line they can curve,
and will all meet at the physical point.


\section{Clover fermions}


The above properties are general; we now apply them to `clover' or 
$O(a)$-improved Wilson fermions. There are now some additional
points to consider when relating the bare quark mass (which is the input
parameter) to the renormalised quark mass. The problem is that for
fermions with no chiral symmetry the singlet, S, and non-singlet,
NS, quark mass can renormalise differently which means that the
relation to the bare quark masses and hence $\kappa$, which is the
adjustable simulation parameter, is more complicated \cite{gockeler04a}
\begin{eqnarray}
   m^{\R}_q &=& Z_m^{\NS}(m_q - \overline{m}) + Z_m^{\Si}\overline{m}
                                                                \nonumber \\
         &=& Z_m^{\NS}(m_q + \alpha_Z\overline{m}) \,,
\label{mr2mbare}
\end{eqnarray}
($q = l$, $s$) with $\alpha_Z = (Z_m^{\Si} - Z_m^{\NS})/ Z_m^{\NS}$ and
bare quark mass defined by
\begin{eqnarray}
   am_q = {1 \over 2} \, 
             \left( { 1 \over \kappa_q} - {1 \over \kappa_{sym;c}} \right) \,,
\end{eqnarray}
where $\kappa_{sym;c}$ is defined by the vanishing of the quark mass
along the symmetric line, i.e.\ for $3$ mass degenerate flavours.
Now from LO $\chi$PT we have
\begin{eqnarray}
   \third (2(am_K)^2 + (am_\pi)^2)
   \propto \twonineth (1 + \alpha_Z) a\overline{m} \,,
\end{eqnarray}
and so the path $a\overline{m} = \mbox{constant}$ remains as
$\third(2(am_K)^2 + (am_\pi)^2)$ constant. This translates to
\begin{eqnarray}
  \kappa_s 
      = { 1 \over { {3 \over \kappa_{sym}^{(0)}} - {2 \over \kappa_l} } } \,,
\label{kappas_given_kappal}
\end{eqnarray}
where $\kappa_{sym}^{(0)}$ is the appropriate $\kappa$ on the
$SU(3)$ flavour symmetric line. As discussed previously higher
order corrections have zero derivative at the flavour $SU(3)$
symmetric point and so should be small.

$O(a)$-improvement also leads to a change in the coupling constant,
$g_0^2 \to \tilde{g}_0^2 = g_0^2(1 + b_g a\overline{m})$,
\cite{luscher05a,bhattacharya05a}. However for our trajectory
as $\overline{m}$ is held constant then $\tilde{g}_0^2$ also remains constant.
Also, \cite{bhattacharya05a}, $\overline{m}^{\R} = Z_m^{\Si} \,
[ 1 + ( 3\overline{d}_m  + d_m u(\xi) ) a\overline{m} ] \overline{m}$
where $\overline{d}_m$, $d_m$ are improvement coefficients and
$u(\xi) = 3 - 4\xi + 2\xi^2$ with $\xi =  m_l / m_{sym}^{(0)}$.
As $u(\xi)$ has a minimum at $\xi = 1$, it is again flat
at the symmetry point, as expected. We have estimated that with
the parameters used below, the correction term in $\overline{m}^R$ remains
very small and so we shall ignore this here.

The particular clover action used here consists of the tree level Symanzik
improved gluon action together with a mild `stout' smeared fermion
action. Further details, \cite{horsley08a,cundy09a}, and a determination
of the non-perturbative, NP, coefficient used for the clover term
are described in \cite{cundy09a}. Simulations have been performed using
the Hybrid Monte Carlo, HMC, algorithm with mass preconditioning for
$2$ mass-degenerate flavours and the rational HMC, \cite{clark06a}
for the $1$-flavour. Two programmes were used, a Fortran programme,
\cite{gockeler07a}, and also the Chroma programme, \cite{edwards04a}.
All the runs described here are on $24^3\times 48$ lattices at
$\beta = 5.50$, $c_{sw}= 2.65$, \cite{cundy09a} (we describe the
determination of the scale later). We first need to determine
the symmetric point, using eq.~(\ref{physical_point}).

\begin{figure}[t]
   \hspace{0.50in}
   \epsfxsize=12.00cm 
      \epsfbox{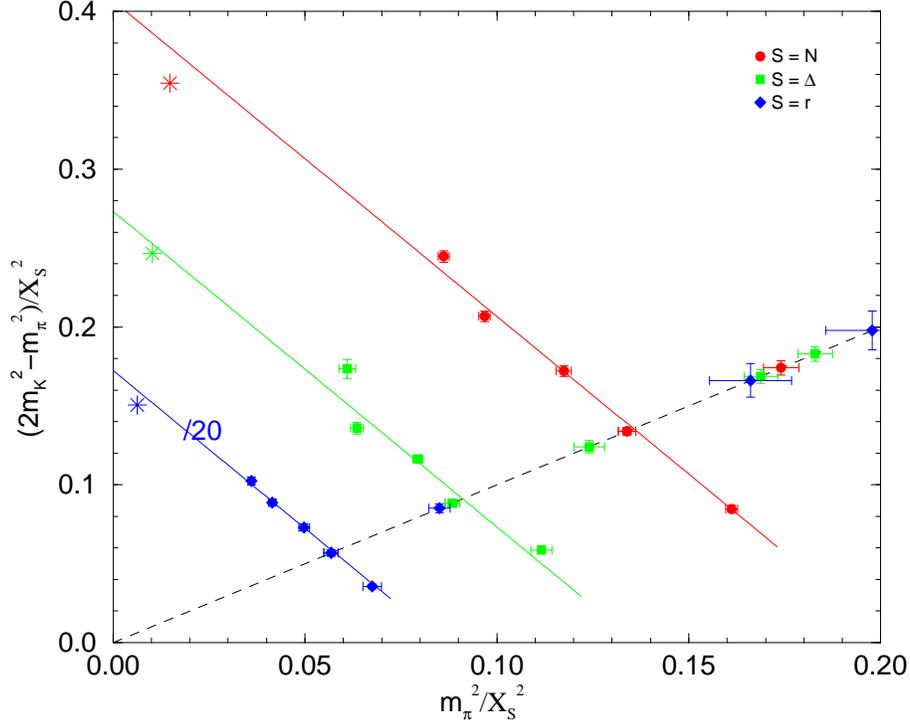}
   \caption{$(2m_K^2-m_\pi^2)/X_S^2$ ($y$-axis) against
            $m_\pi^2/X_S^2$ ($x$-axis) using $S = N$ (red circles)
            $S = \Delta$ (green squares) and $S = r$ (blue diamonds).
            The $SU(3)$ flavour symmetric line ($y = x$) is the dashed line.
            (For convenience the results for $S = r$ have been
            divided by a factor of $20$.)
            The experimental points using the three singlet
            quantities, $X_S$, $S = N$, $\Delta$, $r$
            are shown as stars.
            The solid lines are fits using eq.~(\ref{ds_singlet_fit}).}
\label{b5p50_mps2oX2_2mpsK2-mps2oX2}
\end{figure}

A series of runs on the $SU(3)$ flavour symmetric line gives an estimate
of our starting value $\kappa_{sym}^{(0)}$. Some are shown in
Fig.~\ref{b5p50_mps2oX2_2mpsK2-mps2oX2}
(where we plot $(2m_K^2-m_\pi^2)/X_S^2$ against $m_{\pi}^2/X_S^2$
using the scales $X_S$ with $S = N$, $\Delta$ and $r$) as the points
lying on the $y = x \equiv m_\pi^2/X_S^2$, black dashed line. We seek the
point on the symmetric line where eq.~(\ref{physical_point}) holds. 
This gives an estimate for $\kappa_{sym}^{(0)}$. We find
$\kappa_{sym}^{(0)} = 0.12090$ which we shall take as our starting value.

From this we can now, given a $\kappa_l$, find the corresponding $\kappa_s$
using eq.~(\ref{kappas_given_kappal}). After some experimentation
we chose the $\kappa_l$, $\kappa_s$ values given in Table~\ref{kappa_values}.
\begin{table}[h]
   \begin{center}
      \begin{tabular}{ccc}
         $\kappa_l$ & $\kappa_s$ &               \\
         \hline
         0.12083    & 0.12104    &  $m_l > m_s$  \\
         0.12090    & 0.12090    &  $m_l = m_s$  \\
         0.12095    & 0.12080    &  $m_l < m_s$  \\
         0.12100    & 0.12070    &  $m_l < m_s$  \\
         0.12104    & 0.12062    &  $m_l < m_s$  \\
      \end{tabular}
   \end{center}
\caption{$(\kappa_l, \kappa_s)$ values simulated on $24^3\times 48$
         lattices.}
\label{kappa_values}
\end{table}
Note that it is possible to choose $\kappa_l$, $\kappa_s$ values
(here $(0.12083, 0.12104)$) such that $m_l > m_s$. In  this strange
world, as previously mentioned, we would expect to see an {\it inversion}
of the particle spectrum, with for example the nucleon being the
heaviest octet particle.

These results%
\footnote{Preliminary results were given in \cite{bietenholz09a}.}
are also shown in Fig.~\ref{b5p50_mps2oX2_2mpsK2-mps2oX2}. Each data
set comprises $\sim O(2000)$ trajectories. (Note that our lowest
pion mass has $m_\pi L \sim 3.4$ and may be showing some sign of finite
size effects.) Also shown is a fit to constant $2m_K^2 + m_\pi^2$ by
\begin{equation}
   {2m_K^2 - m_\pi^2 \over X_S^2} = c_S - 2\,{m_\pi^2 \over X_S^2} \,.
\label{ds_singlet_fit}
\end{equation}
The numerically simulated points all lie (approximately) on the
line of constant $2m_K^2 + m_\pi^2$. There is also consistency
between the  various singlet quantities used.

\begin{figure}[t]
   \hspace{0.50in}
   \epsfxsize=12.00cm 
      \epsfbox{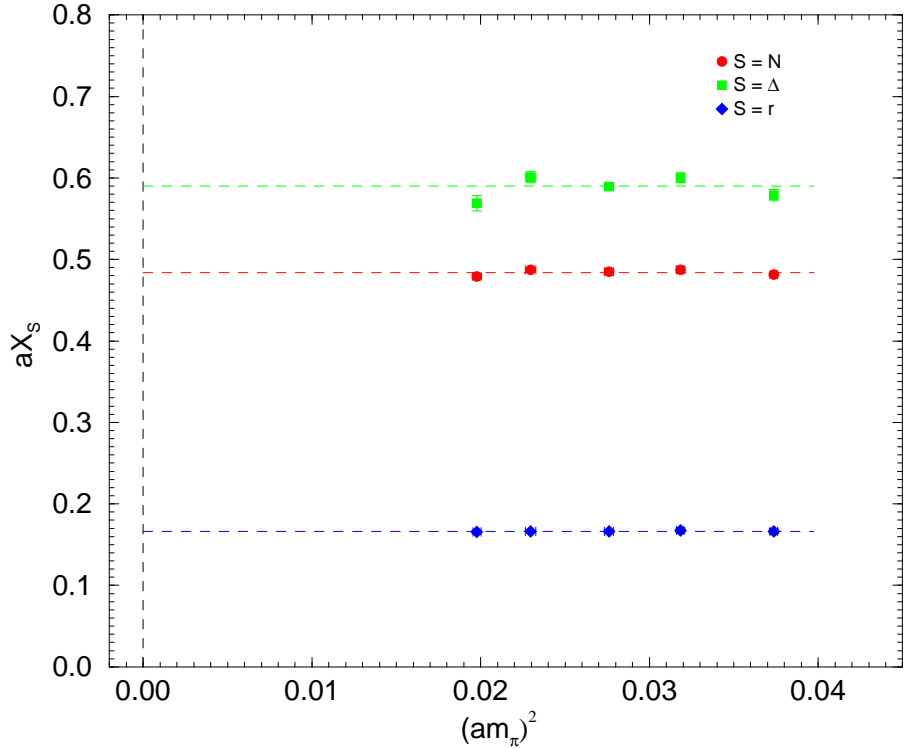}
   \caption{$aX_S$ against $(am_\pi)^2$ for $S = N$ (red circles)
            $S = \Delta$ (green squares) and $S = r$ (blue diamonds).
            Constant fits are also shown (dashed lines).}
\label{b5p50_amps2oX2_aX}
\end{figure}

To determine the scale we again use the constancy of $X_S$.
In Fig.~\ref{b5p50_amps2oX2_aX}
we show $aX_S$ against $(am_\pi)^2$ for $S = N$, $\Delta$, $r$
together with constant fits. (Although the physical limit is also
shown, as we are making a constant fit, this is not important here.)
Then from $aX_S = \mbox{constant}$ we can determine the scale giving
$a = 0.083\,\mbox{fm}$, $0.084\,\mbox{fm}$ for $S = N$, $\Delta$
respectively. We use in future the $S = N$ or $a = 0.083\,\mbox{fm}$
value. This means that the box size, $L \sim 2\mbox{fm}$.
While for $X = r$ the numerical results are the flattest, the physical
value is less well known. However reversing the argument and using
$a = 0.083\,\mbox{fm}$ gives $r_0 = 0.50\,\mbox{fm}$.

As is apparent from Fig.~\ref{b5p50_mps2oX2_2mpsK2-mps2oX2}, we have
slightly underestimated $\kappa_{sym}^{(0)}$ (reflected by the fact
that $\third c_S$ from eq.~(\ref{ds_singlet_fit}) is not quite equal to
$\third(2m_K^2+m_\pi^2)/X_S^2|^*$, cf eq.~(\ref{physical_point})).
We can estimate the significance of this on the kaon mass,
by taking the value of $(2m_K^2-m_\pi^2)/X_N^2$ at the physical
pion mass, $(m_\pi^2/X_N^2)|^*$ in the figure.
This gives $m_K \sim 509\,\mbox{MeV}$, a $\sim 3\%$
discrepancy when compared to the experimental kaon mass of
$\sim 494\,\mbox{MeV}$.


\section{Hadron spectrum}

\begin{figure}[t]
   \hspace{0.50in}
   \epsfxsize=12.00cm 
   \epsfbox{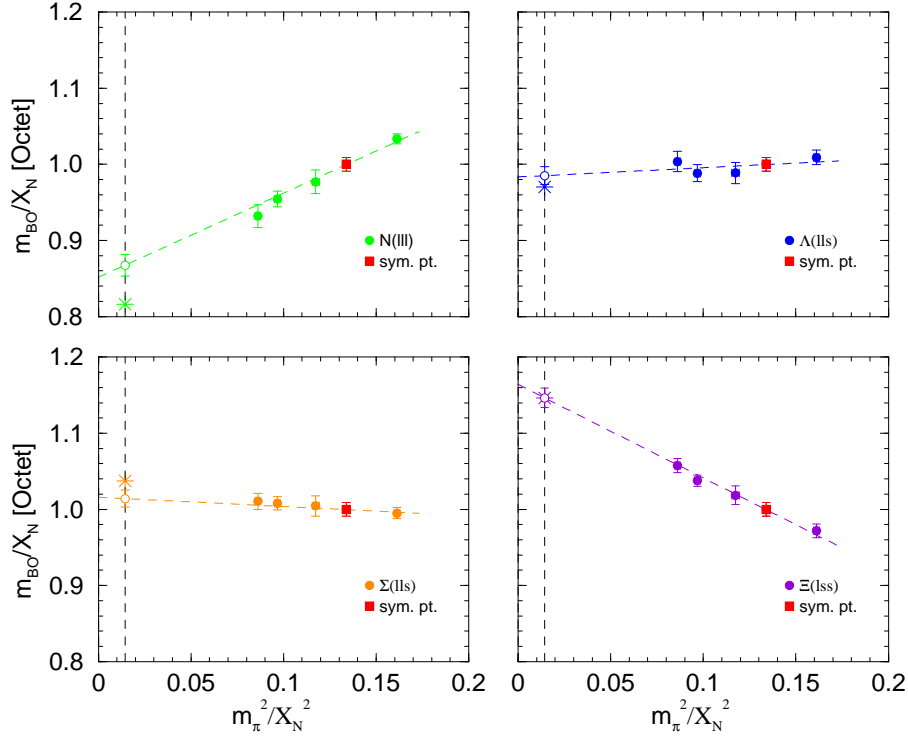}
   \caption{The octet baryon masses $O$ where $O = N$,
            $\Lambda$, $\Sigma$ and $\Xi$ using the scale $S = N$,
            upper left to lower right respectively, together with
            the fit from eq.~(\ref{octet_fit}).
            The common $SU(3)_F$ symmetric quark mass value is
            shown in red. The experimental values are shown
            with stars.}
\label{b5p50_mps2omNOpmSigOpmXiOo32-sym_mNOomNOpmSigOpmXiOo3_100210-jnt_4pic}
\end{figure}
\begin{figure}[h]
   \hspace{0.50in}
   \epsfxsize=12.00cm 
      \epsfbox{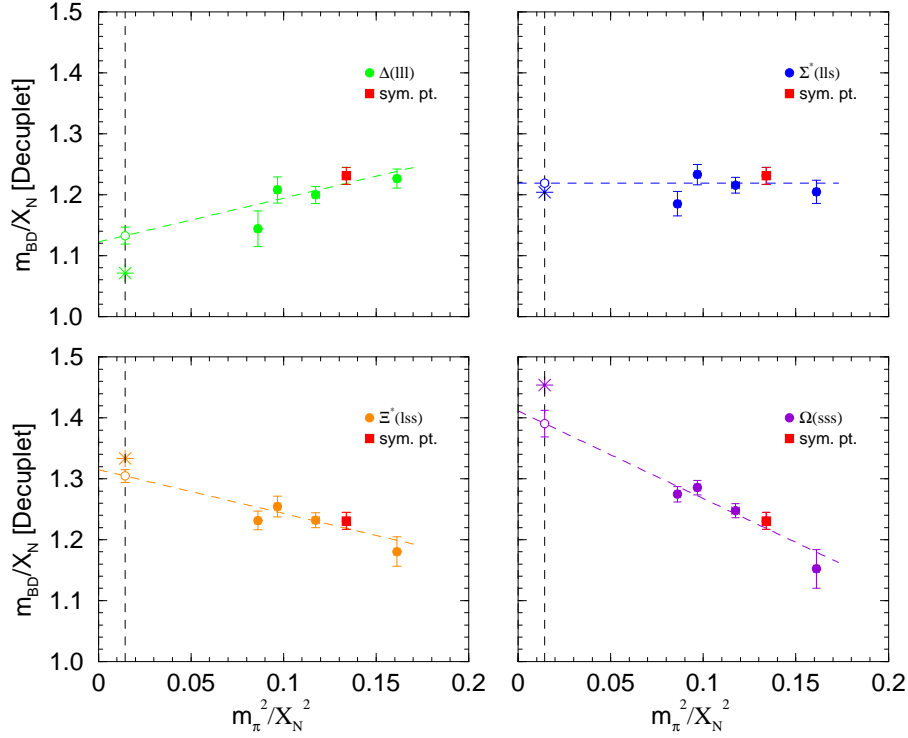}
   \caption{The decuplet baryon masses, $D$, where
            $D = \Delta$, $\Sigma^*$, $\Xi^*$ and $\Omega$ (also using the scale
            $S = N$), together with the fit from eq.~(\ref{decuplet_fit}).
            The common $SU(3)_F$ symmetric quark mass value is
            shown in red. The experimental values are shown
            with stars.}
\label{b5p50_mps2omNOpmSigOpmXiOo32-sym_mDDoX_100210-jnt_4pic}
\end{figure}

The octet (and decuplet) baryon masses are degenerate at
the $SU(3)$ flavour symmetric point and then fan out.
As a first example we now give some mass results in
Fig.~\ref{b5p50_mps2omNOpmSigOpmXiOo32-sym_mNOomNOpmSigOpmXiOo3_100210-jnt_4pic}
for the baryon octet:  $N$, $\Lambda$, $\Sigma$, $\Xi$, together
with a constrained fit from LO $\chi$PT, e.g.\ \cite{walker-loud04a}
about the $SU(3)$ flavour symmetric point,
\begin{eqnarray}
   m_N      &=& A_N + 2(\alpha_N+\beta_N)(m_\pi^2 - m_\pi^2|_{sym}^{(0)})
                                                             \nonumber \\
   m_\Sigma &=& A_N + (\alpha_N-2\beta_N)(m_\pi^2 - m_\pi^2|_{sym}^{(0)})
                                                             \nonumber \\
   m_\Lambda &=& A_N - (\alpha_N-2\beta_N)(m_\pi^2 - m_\pi^2|_{sym}^{(0)})
                                                             \nonumber \\
   m_\Xi   &=& A_N - 3\alpha_N (m_\pi^2 - m_\pi^2|_{sym}^{(0)}) \,,
\label{octet_fit}
\end{eqnarray}
where $A_N \equiv m_{0N}+2(\alpha_N+\beta_N+3\sigma_N)\overline{m}$, $\alpha_N$,
$\beta_N$ are the three fit parameters (in the figure we have also normalised
the masses with $X_N$). Similarly in
Fig.~\ref{b5p50_mps2omNOpmSigOpmXiOo32-sym_mDDoX_100210-jnt_4pic}
results are given for the baryon decuplet:
$\Delta$, $\Sigma^*$, $\Xi^*$, $\Omega$, again together with the
constrained fit from LO $\chi$PT, e.g.\ \cite{tiburzi04a},
\begin{eqnarray}
   m_\Delta    &=& A_\Delta + 2\gamma_\Delta(m_\pi^2 - m_\pi^2|_{sym}^{(0)})
                                                             \nonumber \\
   m_{\Sigma^*} &=& A_\Delta
                                                             \nonumber \\
   m_{\Xi^*}    &=& A_\Delta - 2\gamma_\Delta(m_\pi^2 - m_\pi^2|_{sym}^{(0)})
                                                             \nonumber \\
   m_\Omega    &=& A_\Delta - 4\gamma_\Delta (m_\pi^2 - m_\pi^2|_{sym}^{(0)}) \,,
\label{decuplet_fit}
\end{eqnarray}
where $A_\Delta \equiv m_{0\Delta} + 2(\gamma_\Delta-3\sigma_\Delta)\overline{m}$,
$\gamma_\Delta$ are the two fit parameters.

Both these figures illustrate the `proof of concept'
of the method described here. The correct ordering of the
particle spectrum has been achieved (including the anti-ordering behind
the symmetric point). The masses (using the
scale determined by $S=N$) are given in Table~\ref{baryon_masses}.
\begin{table}[h]
   \begin{center}
      \begin{tabular}{crrr}
         Particle      & Expt. [MeV] & Result [MeV] & Discrep. \\
         \hline
         $m_N$         & 939        & 998(16)     & $6\%$    \\
         $m_\Lambda$    & 1116       & 1133(14)    & $2\%$    \\
         $m_\Sigma$     & 1193       & 1166(13)    & $2\%$    \\ 
         $m_\Xi$        & 1318       & 1318(15)    & $0\%$    \\ 
         \hline
         $m_\Delta$     & 1232       & 1303(16)    & $5\%$    \\
         $m_{\Sigma^*}$  & 1385       & 1402(05)    & $1\%$    \\ 
         $m_{\Xi^*}$     & 1533       & 1501(12)    & $2\%$    \\ 
         $m_\Omega$     &  1673       & 1599(25)    & $4\%$    \\
         \hline
      \end{tabular}
   \end{center}
\caption{Masses for the baryon octet and decuplet.}
\label{baryon_masses}
\end{table}
The results are already within a few percent of their experimental values, and
clearly show that higher orders of $\chi$PT are small.
We can expect an improvement for results closer to the physical pion mass.

\begin{figure}[h]
   \hspace{0.50in}
   \epsfxsize=12.00cm 
      \epsfbox{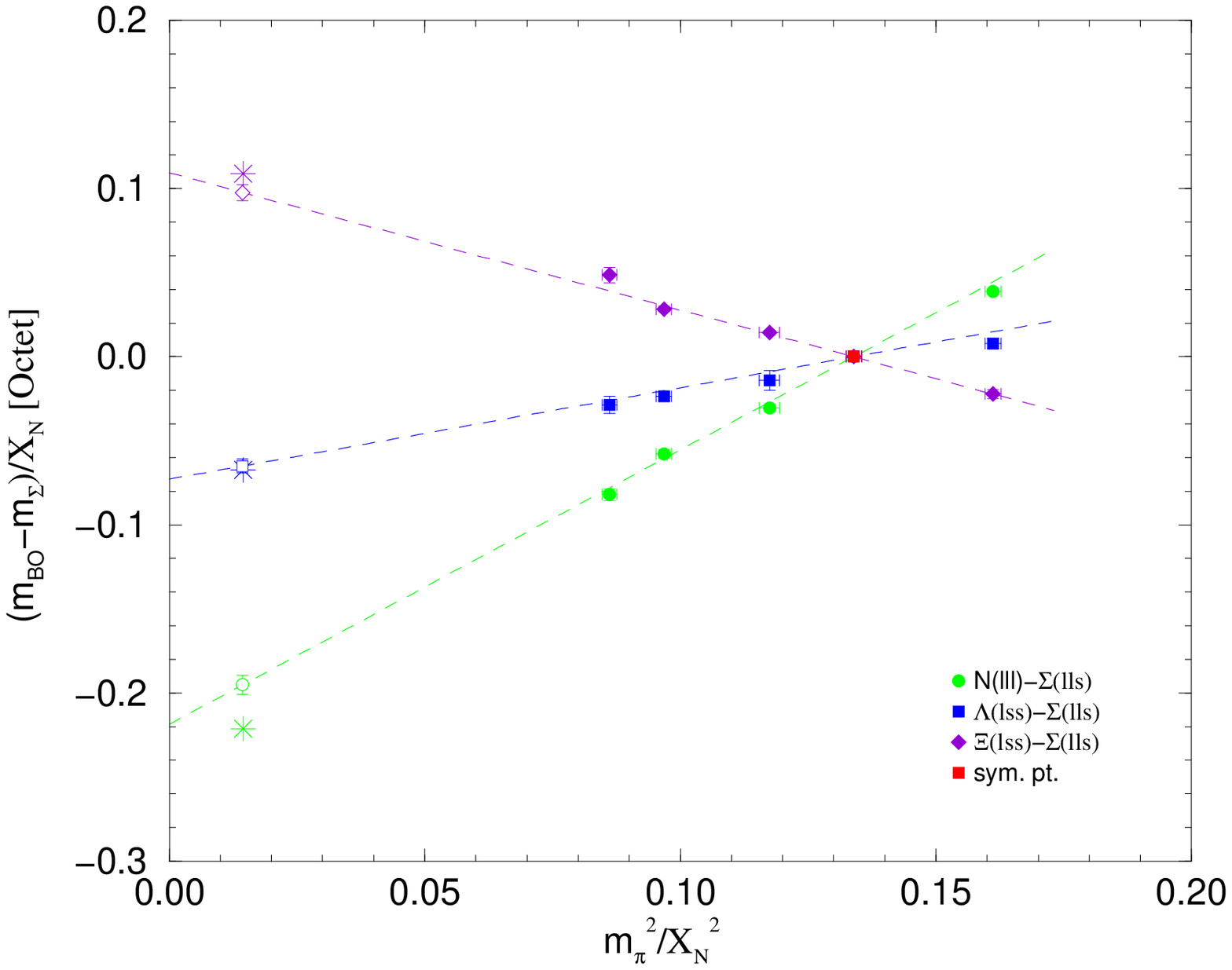}
   \caption{The octet baryon mass splittings $O - \Sigma$ where $O = N$,
            $\Lambda$ and $\Xi$ using the scale $S = N$.
            The common $SU(3)_F$ symmetric quark mass value is
            shown in red. The experimental values are shown
            with stars.}
\label{b5p50_mps2omNOpmSigOpmXiOo32-sym_mNOmmSigOomNOpmSigOpmXiOo3_split-jnt}
\end{figure}

Finally to illustrate better the `fanning' out of the results
we show in
Fig.~\ref{b5p50_mps2omNOpmSigOpmXiOo32-sym_mNOmmSigOomNOpmSigOpmXiOo3_split-jnt}
mass splittings for the baryon octet: $m_N-m_\Sigma$, $m_\Lambda-m_\Sigma$,
$m_\Xi-m_\Sigma$ together with the constrained fit from eq.~(\ref{octet_fit}).
Considering mass splittings has the advantage that the results
can be obtained directly from the ratio of correlation functions,
which leads to a significant reduction in the error.


\section{Conclusions}


We have suggested here that the simplest way of approaching
the physical point in lattice simulations is to hold the singlet
quark mass fixed from a point on the $SU(3)$ flavour symmetric line.
This has been shown both theoretically and numerically 
(using an  NP $O(a)$-improved $2+1$ flavour clover action)
to lead to very smooth results in the extrapolation of singlet
quantities to the physical pion mass. Exploratory results for the
hadron mass spectrum give masses in the correct order
(including {\it inversion} when $m_l > m_s$ i.e.\ we can simulate
a strange world where, for example, the nucleon can decay).
Furthermore the extrapolated masses for both the baryon octet
and decuplet are within a few percent of their experimental values.
To improve the situation further clearly we need simulations closer
to the physical pion mass. At present in our simulations the pion mass
decreases from $\sim 450\,\mbox{MeV}$ to $\sim 335\,\mbox{MeV}$
while the kaon mass increases from $\sim 400\,\mbox{MeV}$ to 
$\sim 450\,\mbox{MeV}$, (experimentally $m_\pi = 138\,\mbox{MeV}$,  
$m_K = 494\,\mbox{MeV}$). Also to improve the accuracy of the
approach to the physical pion mass another line of constant singlet mass
should be found, which will allow interpolation/extrapolation around
the physical pion mass. Finally we note that even LO $\chi$PT seems
to be working very well around the flavour symmetric point.
Further results will be published elsewhere, \cite{bietenholz10a}.


\section*{Acknowledgements}


The numerical calculations have been performed on the IBM
BlueGeneL at EPCC (Edinburgh, UK), the BlueGeneL and P at
NIC (J\"ulich, Germany), the SGI ICE 8200 at HLRN (Berlin-Hannover, Germany)
and the JSCC (Moscow, Russia). We thank all institutions.
The BlueGene codes were optimised using Bagel, \cite{boyle05a}.
This work has been supported in part by the EU Integrated Infrastructure
Initiative {\it Hadron Physics} and by the DFG under contract SFB/TR 55
(Hadron Physics from Lattice QCD).



\end{document}